\documentclass[aps, prd, twocolumn, superscriptaddress, nofootinbib]{revtex4}
\usepackage{graphicx}
\usepackage{dcolumn}
\usepackage{amssymb}
\usepackage{amsmath}
\usepackage{array}
\usepackage{float}

\begin{document}

\newcommand{\Eq}[1]{\mbox{Eq. (\ref{eqn:#1})}}
\newcommand{\Fig}[1]{\mbox{Fig. \ref{fig:#1}}}
\newcommand{\Sec}[1]{\mbox{Sec. \ref{sec:#1}}}

\newcommand{\PHI}{\phi}
\newcommand{\PhiN}{\Phi^{\mathrm{N}}}
\newcommand{\vect}[1]{\mathbf{#1}}
\newcommand{\Del}{\nabla}
\newcommand{\unit}[1]{\;\mathrm{#1}}
\newcommand{\x}{\vect{x}}
\newcommand{\ScS}{\scriptstyle}
\newcommand{\ScScS}{\scriptscriptstyle}
\newcommand{\xplus}[1]{\vect{x}\!\ScScS{+}\!\ScS\vect{#1}}
\newcommand{\xminus}[1]{\vect{x}\!\ScScS{-}\!\ScS\vect{#1}}
\newcommand{\diff}{\mathrm{d}}

\newcommand{\be}{\begin{equation}}
\newcommand{\ee}{\end{equation}}
\newcommand{\bea}{\begin{eqnarray}}
\newcommand{\eea}{\end{eqnarray}}
\newcommand{\vu}{{\mathbf u}}
\newcommand{\ve}{{\mathbf e}}

        \newcommand{\vU}{{\mathbf U}}
        \newcommand{\vN}{{\mathbf N}}
        \newcommand{\vB}{{\mathbf B}}
        \newcommand{\vF}{{\mathbf F}}
        \newcommand{\vD}{{\mathbf D}}
        \newcommand{\vg}{{\mathbf g}}
        \newcommand{\va}{{\mathbf a}}
\newcommand{\mn}{_{\mu\nu}}
\newcommand{\ab}{_{\alpha\beta}}
\newcommand{\uab}{^{\alpha\beta}}
\newcommand{\umn}{^{\mu\nu}}
\newcommand{\dof}{{\it dof} }


\title{Testing General Free Functions in Preferred Scale Theories}

\newcommand{\addressImperial}{Theoretical Physics, Blackett Laboratory, Imperial College, London, SW7 2BZ, United Kingdom}
\newcommand{\addressStMarys}{Applied Physics, St Mary's University, Twickenham, London, TW1 4SX, United Kingdom}

\author{Ali Mozaffari}
\email{ali.mozaffari@imperial.ac.uk}
\affiliation{\addressImperial}

\date{\today}

\begin{abstract}
Building on previous work, we explore the parameter space of general free functions in non-relativistic modified gravity theories motivated by k-essence
and other scalar-tensor theories.  Using a few proposed tests, we aim to update Solar System based constraints on these ideas in line with previous
theories and suggest their utility in constraining modification to GR, potentially
even being able to test k-essence type theories.  
 
\end{abstract}

\keywords{cosmology, modified gravity}
\pacs{04.50.Kd, 04.80.Cc}

\maketitle


\section{Introduction}
Since Einstein's unveiling of his General Theory of Relativity (GR) more
than 100 years ago, it has proven to be a {\it very} resilient idea, surviving tests in the weak, cosmological and strong field regimes with exquisite precision.  There remain however both a number of conceptual and experimental questions that GR has yet to answer, such as how can we marry together gravity with quantum theory and what is the nature of the dark sector.
 In recent years, theories of modified gravity~\cite{Clifton11} have re-emerged as leading contenders to answer some of these open problems.  One such branch of theories involves adding additional \dof's such that dynamics around additional (but perhaps arbitrarily inserted) acceleration scale(s) may vary become important.  These preferred acceleration scale theories first appeared in the guise of MOdified Newtonian Dynamics (MOND)~\cite{Milgrom:1983ca,aqual} and have spawned a series of relativistic extensions~\cite{teves,aether1,MONDreview}.  Such ideas originally appeared as a counterpart to the dark sector, however the two may exist in harmony and perhaps ease the tension on neutrinos as a candidate dark matter particle~\cite{yusaf1,DMastro}.

In this work, we aim initially to abstract ourselves from the details of
any relativistic completion of a preferred scalar theory and instead focus
on computing the consequences in the weak field limit of some general effective theory.  We will be specialising our computations to two specific tests around gravitational saddle points  (SP's) \begin{enumerate}\item{\bf Tidal Stresses - }The expected (close to linear)
tidal stresses of a Newtonian gravitational field may be contrasted with
that of some modified gravitational theory that is in some way ``switched on'' in the low acceleration regime around SP's.    \item{\bf Time Delays - }The experimental gravity favourite with a new twist, the effect on the
stress-energy around some region of modified gravity should have an effect on the observed Shapiro time delay~\cite{CliffordPPN} \end{enumerate}
We then consider a particular realisation of this theory, inspired by k-essence
models and consider prospects for constraints.  

The structure of this paper is as follow: Firstly we recap analytical solutions
for preferred scale theories around SP's making use of parameterisable free
functions to demonstrate effects in different regimes.  We move on to the
effects possible by generalising our free functions to include both potential
$\phi$ and gradient of potential $|\Del\phi|$, considering similar analytical
regimes.  Using first tidal stresses and then time delays as probes, we explore
the different effects on observables with these general functions.  We conclude with some thoughts on constraints for the future and
possibilities for detection.


\section{Analytical Solutions}

\subsection{The $\vU$ Formalism for Saddle Points}
\subsubsection{Recap of the basics}\label{Ubasics}
Our starting point is in the non-relativistic limit of these modified gravity
theories.   A central object here is the (physical) gravitational potential,
 denoted $\Phi$, we define this simply by saying in the relativistic limit
 of this theory, particles feel an acceleration $-\Del\Phi$.  We will stick
 to the Type I,II,III nomenclature (see Appendix \ref{theorytype})$\Phi$ may be decomposed in Type I theories as \bea \Phi = \Phi_N + \phi \eea such that the Newtonian as usual satisfies \bea \Del^2\Phi_N = 4\pi G \rho\eea and the fifth force field $\phi$ has equation of motion \bea \Del\cdot(a_1 \Del\phi) = C_\rho G \rho\eea where $a_1$ is a free function, $C_\rho$ is a dimensionless coupling constant (in these models typically $C_\rho \ll 1$) and $\rho$ is the baryonic matter density.  An appealing suggestion here is to examine these
as effective theories for departures from GR, which we can then be subject to constraints from experiment.  Around SP's, we can take the $\rho \rightarrow 0$ limit and hence \bea \Del\cdot(a_1
\Del\phi) = 0\eea Defining the variable \bea \vU = -a_1\frac{C_\rho}{4\pi}\frac{\Del\phi}{\alpha}\eea
where $\alpha$ is a constant with units of acceleration.  Further to this
we can define the dimensionless scalar \bea z = \frac{C_\rho}{4\pi}\frac{\Del
\phi}{\alpha}\eea Such that \bea U &=& |\vU| = a_1\,z  \eea This redefinition allows us to rewrite (\ref{phieom}) as a system of vector equations \bea \Del\cdot \vU &=& 0 \label{vecteqn1}\\ \Del \wedge \left(\frac{\vU}{a_1}\right) &=& 0 \label{vecteqn2}\eea Expanding
out (\ref{vecteqn2}) and collecting terms together, \bea  a_1 \Del\wedge \vU + \vU \wedge \Del a_1 &=& 0\eea In the case of $a_1 = a_1(z)$, then we can simply rewrite $U = U(z)$ and hence $a_1(z) = a_1(U)$ meaning that  \bea \Del a_1 = \frac{d\,a_1}{d\,U^2}\Del U^2\eea
and hence Equation (\ref{vecteqn2}) becomes \bea M\, U^2\, \Del \wedge \vU + \vU \wedge \Del U^2 &=& 0 \label{vecteqn3} \\  \frac{d \ln U^2}{d \ln a_1} &=& M\eea

We can illustrate the types of solutions in different regimes using a parameterised
free function (previously considered in detail~\cite{aliscaling}), \bea a_1 = \frac{z^a}{(1+z^b)^{a/b}} \label{paramfreefun} \eea but we stress the techniques
described here are applicable to any choice of $a_1$.   

In the proximity of a gravitational saddle point (SP), it is well know that the Newtonian field is linearised.  To a good approximation, we may introduce a truncated multi-pole expansion of the Newtonian field (in spherical polar coordinates, $r,\psi,\theta$) around the SP, \bea \Del \Phi_N &=& -A r \vN \\ \vN  &=& N_r \ve_r + N_\psi \ve_\psi \\ N_r &=& \frac{1}{4}\left(1 + 3 \cos 2\psi\right) \\ N_\psi &=& -\frac{3}{4}\sin 2\psi\eea where $A$ is the expected Newtonian tidal stress and the additional $\theta$ coordinate is absent due to the spherical symmetry present.  We assign the Newtonian contribution
to the field $\vU$, found for $a_1 \rightarrow 1$ as \bea \vU_0 &=& -\left(\frac{C_\rho}{4\pi}\right)^2\frac{\Del\Phi_N}{\alpha}
\eea The utility of this linear
approximation is great, particularly as it sets up a separable ansatz for
solutions in the large and small limits of $U$.  

The boundary between the two regimes in $U$ is found at $|\vU|^2
\simeq 1$.  The intuition here (as justified in~\cite{bekmag}) is that departures from spherical symmetry are subdominant, such that \bea\vU \simeq \vU_0 &\Rightarrow& r^2 \left(\cos^2 \psi + \frac{1}{4}\sin^2 \psi\right) = \left(\frac{4 \pi}{C_\rho}\right)^4 \left(\frac{\alpha}{A}\right)^2 = r_0^2\nonumber \\ \\ \Rightarrow \vU_0 &=& A\left(\frac{r}{r_0}\right)\vN\eea This suggests an ellipsoidal boundary
around the saddle point, with semi-major axes $r_0$ between which there exists
two different regimes.  

It is found in general~\cite{aliscaling} that solutions in the $U \gg 1$ limit take the form \bea \vU &=& \vU_0 + \vU_2 \\ \vU_2 &=& \left(\frac{r}{r_0}\right)^{1-b}\left(F(\psi)\,\ve_r + G(\psi)\, \ve_\psi\right)\eea where $b$ refers to the fall off power of
$z$ in the expansion of $a_1$ in (\ref{paramfreefun}).  We note that this
sets up a perturbative expansion in $\vU$ with \bea \frac{|\vU_2|}{|\vU_0|}\sim
\left(\frac{r}{r_0}\right)^{-b}\eea and suggests that in the large $r$ limit, the fifth force becomes a rescaled Newtonian contribution (effectively renormalising
$G_N$).  In the $U \ll 1$ limit, the behaviour follows \bea \vU = C_{M} \left(\frac{r}{r_0}\right)^{\gamma_a} \left(F_a(\psi)\,\ve_r + G_a(\psi)\, \ve_\psi\right)\label{smallUsoln}\eea where $\gamma_a$ and $C_{M}$ and are constants that can be calculated
depending on the model picked (as detailed in~\cite{aliscaling}) and $a$ is the leading order exponent of $z$ in the expansion of $a_1$ in (\ref{paramfreefun}).
Finally we may attempt to compute the behaviour of $\phi$ field by solving (\ref{vecteqn3}) to find $\vU$ (in the appropriate limit) and then inverting through \bea -\Del\phi = \frac{4 \pi \alpha}{C_\rho}\frac{\vU}{a_1(U)}\eea
and this we may compute both tidal stresses \bea S_{ij} = -\frac{\partial^2
\phi}{\partial x_i \partial x_j}\eea

\subsubsection{More General $a_1(z,\phi)$}
In the more general case of $a_1(z,\phi)$, obviously then $U = U(z,\phi)$.
 But {\it if} we can find $z(U,\phi)$ (which is often possible in particular limits e.g. $z  \ll 1$) then we may write
 $a_1 = a_1(U,\phi)$.  In this way, \bea \Del a_1 = \frac{\partial a_1}{\partial U^2}\Del U^2 +\frac{\partial
 a_1}{\partial \phi} \Del\phi\eea Also since $\vU \propto \Del \phi$ then
 $\vU \wedge \Del \phi = 0$ and so the final system of equations are now mimic the $U(z)$ case of Equation (\ref{vecteqn3}), except that \bea M &=& 1\Biggm/ \frac{\partial \ln a_1}{\partial  \ln U^2} = \frac{2}{1 - a_1\, z_{,U}} \eea where $_{,U} \equiv \frac{\partial}{\partial U}$
Thus for a particular model $P(z,\phi)$, we must first proceed to identify the free function $a_1$, then depending on the limit in consideration expand $a_1(z,\phi)$ as some power series in $z$.  Next compute the relation $U = U(z,\phi)$ and so $z_{,U}$ to find $M$.  


\subsection{Examples}
We can illustrate our techniques using an adaptation of our parameterised
free function,
\bea a_1 &=& \frac{z^a\,u^c}{(1+z^b\,u^{bc/a})^{a/b}}\label{generalfreefun}\\ u &=& \left(\frac{\phi}{v^2}\right)^c \\ M &=& \frac{2}{a}\left(1 + a + z^b
u^{bc/a}\right)\eea where $v$ is some constant with dimension of velocity.  We argue that the value of $v$ should be taken from \bea v^2 \sim A\,r_0^2\eea since it is
of the correct dimension and we see from the case of the Newtonian \bea \frac{\Phi_N}{v^2} = - \left(\frac{r}{r_0}\right)^2 N_r\eea that it sets up a useful cutoff
scale.  This model has the virtue of relatively independently controlled behaviour in each of the $z \ll 1$ and $z \gg 1$ regimes.  We add to this the effect of scaling the modified
potential $\phi$ and seek to explore its behaviour in different regimes around
SP's.

\subsubsection{$z \gg 1$}

In this regime \bea  a_1 &\simeq& \left(1 - \frac{a}{b} \frac{1}{z^b u^{bc/a}}
+ \dots \right)\\ M &\simeq& \frac{2}{a} U^b u^{bc/a}\eea Subsequently (\ref{vecteqn3}) takes the form \bea  \frac{2\, u^{bc/a}}{a}\,U^{b+2} \Del \wedge \vU + \vU \wedge \Del U^2 = 0\eea which may be solved
by assuming the ansatz for $\vU$ \bea \vU = \vU_0 + \vU_2 \eea where $\vU_0$
is again the curl free contribution to the solution and $\vU_2$ is a perturbative contribution with non-zero curl, sourced by $\vU_0$.  If at lowest order
$a_1 \rightarrow$ constant, then we are assured to split up $\vU$ in this
way, however in general a curl free contribution $\vU_0$ may not be present
(c.f. the $z \ll 1$ regime).  In the case of (\ref{generalfreefun}) however we can be satisfied it will be present since $\phi^c\,\Del\phi \propto \Del\left(\phi^{c+1}\right)$
which obviously remains curl free.   

In this regime, the presence of the $\phi$ becomes problematic however our expectation is that we can expand the field as a power series in the large
$r$ limit, \bea \phi \simeq \phi_{\infty} + \epsilon \phi_0 + \epsilon^2\phi_1 + \dots \eea where $\phi_{\infty}$ will be some constant contribution to $\phi$ at $r\rightarrow \infty$.  This means that {\it a priori} we
can find $\phi_0$, \bea \Del\cdot\left(\left(\frac{\phi_0}{v^2}\right)^c \,\Del\phi_0\right) &=& \frac{v^{-2c}}{c+1} \Del^2 (\phi_0)^{c+1} = \frac{C_\rho}{4\pi}\Del^2\Phi_N\nonumber\\ \\ \phi_0 &=& \left(\frac{C_\rho}{4\pi}\frac{|c+1|}{|\Phi_N|^{c}}v^{2c}
\right)^{\frac{1}{c+1}}\Phi_N\eea where the choice of notation
is picked to ensure the correct sign of the force $\vF_\phi$.
 It is required that we know {\it both} of $\vU_0$ and $\phi_0$ in order
 to solve for $\vU_2$ \bea \Del\wedge \vU_2 &=& -\frac{a}{2\,b\,v^{2bc}}\frac{\vU_0 \wedge \Del (U_0)^2}{(U_0)^{b+2}}(\phi_0)^{bc} \eea What becomes obvious
here is that \bea \frac{|\vU_2|}{|\vU_0|}\sim \left(\frac{r}{r_0}\right)^{b(c-1)/(c+1)}\eea which suggests only values of $c$ satisfying $-1 < c \leq1$ will permit perturbative solutions
with $\vU_2$ subdominant to $\vU_0$  as before.  For values of $c$ outside of this, the dominant contribution in expanding $\vU$ will now be $\vU_2$.  We stress that to avoid violations of Solar System constraints, the fifth force can only follow certain limiting behaviour:\begin{itemize} \item{$\Phi$
mimics the Newtonian potential with an appropriately small scaling, such that $G_N$ is effectively renormalised (within limits, such as BBN and the CMB)} \item {$\Del \phi$ becomes subdominant in the large $r$ limit, such that the inner bubble is effectively ``screened''.}\end{itemize} 

Using the notation \bea \vU_2 = U_r \ve_r + U_\psi \ve_\psi \eea we see that (\ref{vecteqn1}, \ref{vecteqn2}) reduce to the coupled ODEs, \bea \frac{1}{r^2}\frac{\partial}{\partial r}(r^2
\,U_r) + \frac{1}{r\sin\psi}\frac{\partial}{\partial \psi}(\sin \psi \,U_\psi)
&=& 0 \\ \left[\frac{\partial}{\partial r}(r\,U_\psi ) - \frac{\partial U_r }{\partial \psi}\right]&=& \frac{s_{b,c}(\psi)}{r^{n-1}} \eea \bea s_{b,c} &=& -\frac{3a}{b}\frac{2^{3b\left(1/2-c\right)}(1+3\cos 2\psi)^{b c} \sin 2\psi}{(5+3\cos 2\psi)^{1 + b/2}} \\ n &=& b\left(1 - \frac{2c}{c+1}\right) \eea which suggests a separable ansatz \bea \vU_2 &=& \left(\frac{r_0}{r}\right)^{n-1}
 \left(\frac{C_\rho}{4\pi}|c+1|\right)^{bc/(c+1)}\vB(\psi)\label{U2general} \\ \vB(\psi) &=& F_{b,c}(\psi)\,\ve_r + G_{b,c}(\psi)\,\ve_\psi \eea where $F_{b,c}, \,G_{b,c}$ satisfy \bea F_{b,c}\,(n-2)(n-3) \nonumber\\  {F_{b,c}}' \cot \psi + {F_{b,c}}'' &=& -({s_{b,c}}' + s_{b,c} \cot \psi) \\ (2-n)\,G_{b,c} &=& s_{b,c} + F_{b,c}' \eea The relative depreciation (for $c > 0 $) or amplification of signal (for $c < 0$) is evident in the second factor of Equation (\ref{U2general}).  Once we characterise the $\vU$ we may find the total contribution of the scalar \dof \bea -\Del \phi &\simeq& \frac{4\pi \alpha}{C_\rho} \left(\frac{v^2}{\phi}\right)^c \vU \left(1 + \frac{a}{b}\left(\frac{\phi}{v^2}\right)^{bc}\frac{1}{U^b} + \dots\right) \nonumber\\ \eea  Naturally we can use the form of the $\vU$
to find the separable
 form of the anomalous force, but for brevity we leave this for Appendix \ref{fullsolns}.  We see from the form
of Equation (\ref{genfifthforce}) that a much richer structure is present
for models with $c\neq0$.  Depending on the model parameters, $(C_\rho,c)$
the additional exterior bubble force may become significantly smaller than the Newtonian background within a short distance.  

By way of a an extreme example, for $c = 1$ (and so $n=0$),
the additional force will have no $r$ dependence (recall this expansion is
valid for $r/r_0 \gg 1$ and so at leading order, \bea \frac{|\Del\phi|}{|\Del\Phi_N|}
\simeq \frac{r_0}{r}\left(\frac{4\pi}{C_\rho\,N_r}\right)^{1/2} + \dots\eea 
Likewise for $c \simeq 0$ \bea \frac{|\Del\phi|}{|\Del\Phi_N|}
\simeq \left(\frac{8\pi}{C_\rho N_r}\right)^c + \dots\eea \newline

\subsubsection{$z \ll 1$}
In this inner bubble regime \bea a_1 &\simeq& \left(\frac{\phi}{v^2}\right)^c z^a \\ U
&\simeq& \left(\frac{\phi}{v^2}\right)^c z^{a+1}\\ 4m
&\rightarrow& \frac{2(a+1)}{a} \eea which can be solved with the $U \ll 1$
solutions of the $c=0$ case.  Putting this together with our ansatz for $\vU$
in (\ref{smallUsoln})
\bea -\Del \phi &\simeq& \frac{4 \pi \alpha}{C_\rho}\frac{\vU}{z^a} \left(\frac{v^2}{\phi}\right)^c \simeq
\vU\left(\frac{v^{2c}}{U^a\,\phi^c}\right)^{1/(a+1)}\\
 -\Del\phi^{a_c/(a+1)} &\simeq& \frac{4 \pi \alpha\,\vU}{C_\rho} \left(\frac{a_c}{a+1}\right)\left(\frac{v^{2c}}{U^a}\right)^{1/(a+1)}  \eea and again we leave the full separable form of the anomalous force to
 Appendix \ref{fullsolns}
 
 \bea H(\psi) &=& \left(\frac{F_a}{({F_a}^2+{G_a}^2)^{1/2}}\right)^{a/a_c}  \\\nonumber a_c &=& a+c+1 \\\nonumber a_\gamma &=& a+\gamma+1\eea  

The condition for divergent tidal stresses becomes, \bea\frac{\gamma-c}{a_c} < 1 \Longrightarrow c > \frac{\gamma-a-1}{2} \label{divclimit} \eea

\subsubsection{An Example}

To clarify matters, we will specialise to a model with $a = 1, \,b = 2$~\cite{aliscaling},
which for the inner bubble presents solutions of the form \bea \vU &=& C_{M}\left(\frac{r}{r_0}\right)^{\gamma} \left(F_1(\psi)\,\ve_r + G_1(\psi)\, \ve_\psi\right) \nonumber\\ F_1 &\simeq& 0.2442 + 0.7246 \cos 2\psi  + 0.0472 \cos 4\psi \nonumber\\ G_1 &\simeq& -0.8334 \sin 2\psi  -0.0368 \sin 4\psi \nonumber\\ C_{M} &\simeq& 1.3163
\nonumber \\ \gamma &\simeq& 1.5256\eea Thus in this model, divergent tidal stress are found for $c > -0.237$.  On the other hand investigating anomalous time delays requires \bea \frac{a_\gamma}{a_c} < 0 \eea This give rise to two branches of solutions \bea \gamma < -(a+1)
&\Longleftrightarrow& c > -(a+1) \\ \gamma > -(a+1) &\Longleftrightarrow& c < -(a+1)\eea 
We see therefore that divergent solutions are feasible in both branches,
contrast to just the $a < 0$ solutions predicted for $c = 0$.

\section{Type II \& III Theories}
Ultimately we may consider other types of non-relativistic effective theory
with these more general free functions, particular those characterised as
types II and III.  Type II theories follow relations of the form \bea \Del^2\phi
= \frac{C_\rho}{4\pi}\Del\cdot(b_1 \Del\Phi_N) \eea where once again $C_\rho$
is some coupling to matter and $a_2$ is a free function.  Typically $b_1$
have been chosen as function of the Newtonian acceleration $|\Del\Phi_N|$ however here we will attempt to relax such a condition to \bea b_1 = b_1(\Phi_N,|\Del\Phi_N|)\eea
By employing the Newtonian linear approximation and a choice of parameterised
free function for illustration, we aim to find suggestive analytical solutions
in this case\bea b_1 &=& \left(1 + \frac{u^c}{w^b}\right)^{a/b}
\\ w &=& \left(\frac{C_\rho}{4\pi}\right)^2\frac{|\Del\Phi|}{\alpha}
= \frac{r}{r_0}\vN \\ u &=& \frac{|\Phi_N|}{v^2} = \frac{1}{2}\left(\frac{r}{r_0}\right)^2 N_r  \eea where the $c=0$ case results in a typical Type II parametrised
free function.  The solutions are best expressed here in terms of the ratio
$r/r_0$ which helps to distinguish inner and outer bubble solutions \bea \Del^2\phi &=& a\, \nu^{1-b/a}\, A \left(\frac{r}{r_0} \right)\,\frac{u^c}{w^{b+1}}f(N,N_r,N_\psi)\\
f &=& N
N_r + N\psi N' - 2\frac{c}{b}\frac{N}{N_r}\left(N_r^2 + \frac{N_\psi
N'}{2}\right)\eea

Likewise Type III theories can be viewed as a recasting Type I theories,
with the caveat that the free function depends on just the physical potential
$\Del\Phi$, \bea c_1 = c_1(|\Del\Phi|) \eea and the effective coupling takes the value $C_\rho = 4\pi$, \bea \Del\cdot(c_1 \Del \Phi) = \Del^2\Phi_N\eea
where we are positing that \bea c_1 = c_1(\Phi,|\Del\Phi|)\eea

This would produce similar results as the type I theory, however we are restricted
to models which have \bea \lim_{|\Del\Phi|\gg a_{pref}}  c_1 &\rightarrow&
C_1 + C_2(\Phi,|\Del\Phi|) + \dots \eea where $C_1$ is a constant close to
unity.  Such a form is required to ensure Solar System tests are not violated.

\section{A Toy Model}
Given that we may consider free functions of the form outlined in Section
??, it would be interested to understand how to motivate them from the point
of view of a relativistic gravitational theory.  Consider first the action for k-essence theories minimally coupled to gravity
\bea S = \int \left(\frac{M^2_{pl}}{2} R + P(X,\phi) - V(\phi)\right)\,\sqrt{-g}\,d^4
x \label{kessenceaction}\eea where $X = -\frac{1}{2}\partial_\mu\phi \partial_\mu \phi g\umn$ is
the canonical kinetic term and obviously $P = X$ in the case of a simple
canonical scalar field.  Different cases of the forms of $P$ are enumerated in the literature and these have been proposed as candidate theories for both inflation and latterly dark energy\footnote{Although there exists a no-go theorem for many classes of these theories as a DE candidate, see~\cite{kessenceDEnogo}}
(DE).  If we consider an inflationary theory, where the choice of
$V(\phi)$ will enforce a regime of a slow-rolling $\phi$ leading to acceleration
expansion of the universe and then as $V(\phi)$ settles into a minima, inflation
ends.  

We can contrast this with a scalar preferred acceleration scale theory~\cite{prefscale}, \bea
S &=& \int\frac{M^2_{pl}}{2} R \,\sqrt{-g}\,d^4 x \nonumber \\ &+& \int\frac{1}{\kappa}\left(\frac{a_1}{\ell}X - V(\phi) - \alpha^2 F(a_1)\right)\,\sqrt{-g}\,d^4x \nonumber \\ &+& \int \mathcal{L}_M\, \sqrt{-\tilde{g}}\,d^4 x \\ \tilde{g}\mn &=& A(\phi) g\mn\eea examining the equations of motion, both for the metric \bea G\mn = 8\pi G \,{T^M}\mn + \frac{16 \pi G}{\kappa} {T^\phi}\mn\eea where
${T^M}\mn$ is the (Einstein frame) matter stress energy and ${T^\phi}\mn$
is the scalar field stress energy.  Additionally for the scalar field, \bea \Del_\mu (a_1 \Del_\nu \phi)
g\umn &=& \kappa \ell(V_{,\phi} + A_{,\phi} \,A \,\mathcal{L}_M) \label{phieom}\\ F_{,a_1} &=& \frac{X}{\alpha^2 \ell} \eea where $_{,\phi} \equiv \frac{\partial}{\partial \phi}$ etc.  This set up ensures that for some free function $a_1$,\bea F_{,a_1}
= G(a_1) = \frac{X}{\alpha^2\ell} \Rightarrow a_1 = {G^{-1}}\left(\frac{X}{\alpha^2\ell} \right)\eea where $G^{-1}$ denotes the inverse function of $G$ and here enforces
 $a_1$ to be a function of just $|\Del\phi|$.  However we are at liberty
to drop this condition and just take the necessary equations of motion from
an action of the form (\ref{kessenceaction}).  Thus we set up a hybrid action,
\bea S &=& \int \left(\frac{M_{pl}}{2} R + \frac{1}{\kappa}\left[P(X,\phi) - V(\phi)\right]\right)\,\sqrt{-g}\,d^4
x \nonumber \\&+& \int \mathcal{L}_m\,\sqrt{-\tilde{g}}\,d^4x\eea We draw
the reader to the similarities between this action and that of Chameleon
theories~\cite{Chameleon}, where $P(X,\phi) = X$ and here we present this theory as
a simple extension.  Making the association in the equations of motion \bea \frac{P(X,\phi)}{X} \longrightarrow \frac{a_1(X,\phi)}{\ell} \eea produces a free function that is now a function of both $X = -\frac{1}{2}|\Del \phi|^2$ and $\phi$.  Likewise we can attempt to expand these perturbatively
in $X$.

Cosmological solutions in the inflationary regime obviously neglect $\mathcal{L}_m$, whilst the spectrum of solutions in the weak-field limit will be regulated by the choice of $P(X,\phi)$ and the magnitude of the effective model parameters.
  Taking the equation of motion (\ref{phieom}) in the weak field limit, assuming a pressureless matter stress energy and the quasi-static limit, \bea {T^\mu}_\mu &=& \rho \\ g\mn &\rightarrow& \eta\mn \\ |\phi| &\ll& 1 \eea\newline 
Additionally we pick a representative conformal factor for illustration, \bea A = e^{m\phi}\eea and assume that since inflation has ended $V_{,\phi} \rightarrow 0$, thus the effective weak field equations of motion become \bea \Del\cdot(a_1\Del\phi)
= \kappa\ell m\, \rho = \frac{\kappa \ell m}{4\pi G} \Del^2 \Phi_N\eea Setting
up a dictionary to bridge these notions, we find \bea C_\rho &=&\kappa \ell
m  \\ X &=& -\frac{z^2}{2}\left(\frac{4\pi\,\alpha}{C_\rho}\right)^2
= -z^2\,X_0 \\ U &=& a_1\sqrt{-\frac{X}{X_0}}=a_1\sqrt{\tilde{X}}\eea
 where  $\tilde{X}$ is a dimensionless counterpart for $X$.   Likewise 
\bea 4m &=& 1\Biggm/ \frac{\partial \ln a_1}{\partial  \ln U^2} = \frac{2}{1 - a_1^2 \tilde{X}_{,U^2}} \eea


\subsection{An Example From k-essence}
We can illustrate our techniques with a example inspired by k-essence~\cite{kessencemodel}, \bea P = \frac{C_1\sqrt{1 + X} - C_2}{\phi^2} + \dots\eea where $C_1,C_2$ are constants and $P$ contains higher order terms that will be subdominant here.  Firstly identifying
\bea a_1 \simeq \frac{C_2 - C_1\sqrt{1-\tilde{X}}}{\tilde{X}\,\phi^2}\eea

and then examining the near SP limit, $\tilde{|X|} \ll 1$
\bea a_1 &\simeq& \frac{C_2-C_1}{\tilde{X}\phi^2}\left(1 + \frac{C_1}{C_2-C_1}\frac{\tilde{X}}{2}
+ \dots \right) \\ U^2 &\simeq& \frac{(C_2-C_1)^2}{\tilde{X}\phi^4} + \dots\\ 4m
&\rightarrow& 1\eea

These results suggest that in this model, we transition from one ``inner bubble'' like regime (as seen in (\ref{smallUsoln})) to another.  This means that in these models, provided the outer bubble effects are consistently
screened, these models would survive Solar System tests, as well as appearing
in exotic inflationary or DE theories.

\subsection{More General $P(X,\phi)$}
Returning to the different forms of generalised pressure, we find that in
each case, a simple scheme for predicting effects can be found \begin{itemize}

\item{\bf Purely Kinetic Function} $P = P(X)$, this case is simply a restating
of a preferred scale model such that \bea a_1(\tilde{X}) = \frac{P(X)}{X}
= \tilde{P}(\tilde{X})\eea which
easily allows analysis using the techniques of Section \ref{Ubasics}.

\item{\bf General Mixed Function} $P = P(X,\phi)$, this case suggests \bea
a_1(\tilde{X},\phi) &=& \frac{P(X,\phi)}{X} = \tilde{P}(\tilde{X},\phi) \eea
Depending on the form of $\tilde{P}$ and regime in question, we are left with differing forms of fifth force which may be perturbative (with $\vU = \vU_0 + \vU_2$) or non-perturbative (with just $\vU$).  
\bea a_1 &\simeq& \tilde{P}_0(\phi) + \tilde{P}_n(\phi)\,\tilde{X}^n + \dots \\  U^2 &\simeq& \tilde{X}\,\left(\tilde{P}_0 + \tilde{P}_n\, \tilde{X}^n + \dots\right)^2 \\4m &\simeq& \frac{2n+1}{n} + \frac{\tilde{P}_0}{n\, \tilde{P}_n \,\tilde{X}^n} + \dots \eea 

\begin{enumerate}\item{\bf
$\tilde{P}_0 \neq 0$ case}, \bea \bar{P}_0(\phi_0)& = &\,\int\,\tilde{P}_0(\phi_0)\,d\phi_0
= \frac{C_\rho}{4\pi}\Phi_N \\ \phi_0 &=& {\bar{P}_0\,}^{-1}(\Phi_N) \\ -\Del\phi &\simeq& \frac{4 \pi \alpha}{C_\rho}\frac{{\tilde{P}_0(\phi_0)}^{2n}}{{\,\tilde{P}_0}(\phi_0)^{2n+1} + \tilde{P}_n(\phi_0) \,U^{2n}}\,\vU \\ \vU &=& \vU_0 + \vU_2 \\ \Del \wedge \vU_2 &=&  -\frac{n\tilde{P}_n(\phi_0)}{\tilde{P}_0(\phi_0)}\frac{\vU_0 \wedge \Del (U_0)^2}{(U_0)^{2-2n}}\eea 
\item{\bf $\tilde{P_0}=0$ case}, \bea \bar{P_n} &=& \int{{\tilde{P}_n\,(\phi)\,}^{\frac{1}{2n+1}}
d\phi}  \\  \vU &=& C_{M}\left(\frac{r}{r_0}\right)^{\gamma_n} \left(F_n \ve_r
+ G_n \ve_\psi\right) \\-\Del\bar{P}_n &\simeq& \frac{4 \pi \alpha}{C_\rho}  \frac{\vU}{ {U\,}^{2n/(2n+1)}} \\ \bar{P}_n &=& \frac{4 \pi \alpha}{C_\rho\,n_\gamma}
\frac{F_n(\psi)\,r^{n_\gamma}}{({F_n}^2 +  {G_n}^2)^{1/2}} {C_M}^{1/(2n+1)}\\ n_\gamma &=& \frac{2n+\gamma_n+1}{2n+1} \nonumber
\\ \phi &=& {\bar{P}\,}^{-1}(r,\psi)\eea
\end{enumerate} where the value of $4m\,(n)$ will determine the parameter $\gamma_n$, $C_M$ (which is usually an $\mathcal{O}(1)$ contribution) and the profile functions $F_n,\,G_n$.

\item{\bf Separable Function} $P = f(\phi)\,g(X)$, this case suggests \bea
a_1(\tilde{X},\phi) = f(\phi)\,\frac{g(X)}{X} = f(\phi)\,\tilde{g}(\tilde{X}) \eea and so a background level calculation must be performed for each regime.
 In each case the leading order expansion in the requisite regimes may take
 two possible forms, \bea a_1
&\simeq& f(\phi)\,\left(\tilde{g}_0 + \tilde{g}_n \tilde{X}^n + \dots\right) \\  U^2 &\simeq& f^2(\phi)\,\tilde{X}\,\left(\tilde{g}_0 + \tilde{g}_n\, \tilde{X}^n + \dots\right)^2 \\ 4m &\simeq& \frac{2n+1}{n} + \frac{\tilde{g}_0}{n\, \tilde{g}_n \,\tilde{X}^n} + \dots \eea We see therefore that this case is
just a reduction of the general mixed function case with \bea \tilde{P}_0
&\equiv& f(\phi)\,\tilde{g}_0 \nonumber\\  \tilde{P}_n &\equiv& f(\phi)\,\tilde{g}_n\eea\\


\end{itemize}

\section{Conclusions}
To conclude, we have considered adaptations to the free functions in different
variations of preferred acceleration modified gravity theories.  Throughout
we have set about to show that current techniques for characterising experimental
observations can be extended to provide a series of concrete predictions
for these theories.  Centering on the low acceleration regions around gravitational
saddle points, we have demonstrated that both divergent tidal stresses and
anomalous time delays provide different ways to constrain these models. 
Bringing about a correspondence between these theories and those arising
in k-essence theories, we posit that a hybrid action (with similarities with
both scalar-tensor ideas e.g. chameleon theories) could explain both preferred
scale behaviour and inflationary dynamics.  Further we argue that these theories
lie naturally within the framework for screened theories, with the anomalous
stresses for such models potentially falling off very quickly outside of
the SP bubble region, further illustrating why such behaviour has not been
noticed previously.

\begin{acknowledgments}
The author is grateful for financial support and hospitality from St Mary's
University, the European Space Agency and particularly Michele Armano during the conception and preparation of this work.    
\end{acknowledgments}

\appendix
\section{Non-Relativistic Limits}\label{theorytype}
Preferred acceleration theories can be classified according to the differences
in their equations of motion.  We assume that the Newtonian potential satisfies
the usual Poisson relation \bea \Del^2\Phi_N = 4\pi G \rho\eea and then additionally
in each theory the gravitational potential follows the relations given in
Table \ref{table:theorytypes}.  Broadly speaking the different non-relativistic
limits can produce additional effects with the observed gravitational constant.
 In theories with two potentials, in the large acceleration limit \bea \eea

\begin{table}[h!]
\centering 
\begin{tabular}{c c c c c cc} 
Name && Potential && Equation of Motion && Free Function\\ [0.5ex] 
\hline\\[0.25ex] 
Type 1 && $\Phi = \Phi_N + \phi$ && $\Del\cdot(a_1\Del\phi) = \frac{C_\rho}{4\pi}\Del^2\phi_N$
&& $a_1(|\Del\phi|)$\\ [2ex]
Type 2A && $\Phi = \Phi_N + \phi$ && $\Del^2\phi = \frac{C_\rho}{4\pi}\Del\cdot(b_1\Del\Phi_N)$&& $b_1(|\Del\Phi_N|)$ \\[2ex]
Type 2B && $\Phi$ && $\Del^2\Phi = \Del\cdot(b_2\Del\Phi_N)$&& $b_2(|\Del\Phi_N|)$ \\[2ex]
Type 3 && $\Phi$ && $\Del\cdot(c_1\Del\Phi) = \Del^2\Phi_N$ && $c_1(|\Del\Phi|)$\\ [2ex] 
\hline 
\end{tabular}
\caption{Summary of non-relativistic limits}\label{table:theorytypes}
\end{table}
\newpage

\section{Full expressions for anomalous forces}\label{fullsolns}
\begin{widetext}\bea -\Del\phi &\simeq& \frac{4 \pi\alpha}{C_\rho} \left(\frac{4 \pi}{C_\rho|c+1|}\frac{2}{N_r}\right)^{\frac{c}{c+1}}\left(\frac{r_0}{r}\right)^{\frac{c-1}{c+1}}  \left[\vN + \left(\frac{C_\rho\,|c+1|}{4 \pi}\right)^{\frac{bc}{c+1}}\left(\frac{r_0}{r}\right)^{n}
\left(\vB - \frac{a\,\vN}{b\,|\vN|^b}\left(\frac{N_r}{2}\right)^{\frac{bc}{c+1}} \right)\right]+\dots \label{genfifthforce}\eea \end{widetext}

\begin{widetext}\bea -\Del\phi
&=& \alpha\,(C_M)^{a_c}\left(\frac{4 \pi}{C_\rho}\right)^{1+\frac{c}{a_c}} \left(\frac{a_c}{a_\gamma}\right)^{1 - \frac{c}{a_c}} \left(\frac{r}{r_0}\right)^{(\gamma-c)/a_c}\left[\frac{a_\gamma}{a_c}H(\psi)\ve_r + H'(\psi)\ve_\psi\right]\eea \end{widetext}

\bibliography{references}

\end{document}